\pdfoutput=1
\documentclass[review]{elsarticle}

\usepackage{lineno,hyperref}
\modulolinenumbers[1]
\usepackage{epsfig}
\usepackage{amsmath}
\usepackage{booktabs}
\usepackage{hyperref}
\UseRawInputEncoding
\usepackage{gensymb}
\usepackage{amssymb}
\usepackage{multirow}
\usepackage{xcolor}

\journal{Journal of \LaTeX\ Templates}

\begin{document}

\begin{frontmatter}

\title{Numerical Study on Beam-based Alignment of SXFEL Undulator Lattice}
%\tnotetext[mytitlenote]{Fully documented templates are available in the elsarticle package on \href{http://www.ctan.org/tex-archive/macros/latex/contrib/elsarticle}{CTAN}.}

%% Group authors per affiliation:
\author[1,2]{Liang Xu}

\author[2,3]{Nanshun Huang}

\author[1]{Qingmin Zhang\corref{cor1}}
\ead{zhangqingmin@mail.xjtu.edu.cn}
\cortext[cor1]{Corresponding author}

\author[4]{Duan Gu\corref{cor2}}
\ead{guduan@sinap.ac.cn}
\cortext[cor2]{Corresponding author}

\author[4]{Haixiao Deng}

\address[1]{Department of Nuclear Science and Technology, School of Energy and Power Engineering, Xi'an Jiaotong University, Xi'an 710049, China}

\address[2]{Shanghai Institute of Applied Physics, Chinese Academy of Sciences, Shanghai 201800, China}

\address[3]{University of Chinese Academy of Sciences, Beijing 100049, China}

\address[4]{Shanghai Advanced Research Institute, Chinese Academy of Sciences, Shanghai 201204 , China}
%\author{Elsevier\fnref{myfootnote}}
%\address{Radarweg 29, Amsterdam}
%\fntext[myfootnote]{Since 1880.}
%
%%% or include affiliations in footnotes:
%\author[mymainaddress,mysecondaryaddress]{Elsevier Inc}
%\ead[url]{www.elsevier.com}
%
%\author[mysecondaryaddress]{Global Customer Service\corref{mycorrespondingauthor}}
%\cortext[mycorrespondingauthor]{Corresponding author}
%\ead{support@elsevier.com}
%
%\address[mymainaddress]{1600 John F Kennedy Boulevard, Philadelphia}
%\address[mysecondaryaddress]{360 Park Avenue South, New York}

\begin{abstract}
The undulator line of the Shanghai soft X-ray Free-electron Laser facility (SXFEL) has very tight tolerances on the straightness of the electron beam trajectory. However, the beam trajectory cannot meet the lasing requirements due to the influence of beam position, launch angle and quadrupole offsets. Traditional mechanical alignment can only control the rms of offsets to about 100 $\mu$m, which is far from reaching the requirement. Further orbit correction can be achieved by beam-based alignment (BBA) method based on electron energy variations. K modulation is used to determine whether the beam passes through the quadrupole magnetic center, and the Dispersion-Free Steering (DFS) method is used to calculate the offsets of quadrupole and BPM. In this paper, a detailed result of simulation is presented which demonstrates that the beam trajectory with rms and standard deviation ($\sigma$) less than 10 $\mu$m can be obtained.
\end{abstract}

%\begin{highlights}
% \begin{itemize}
% 	\item A digital LLRF control system with a generator driven resonant mode and a self-excited loop mode was successfully developed to operate a 1.3 GHz continuous-wave superconducting RF cavity and prevent the occurrence of the ponderomotive instabilities.
% 	\item A digital 1.3 GHz RF cavity simulator was also developed to easily verify the designed algorithms of the LLRF system.
% \end{itemize}
%\end{highlights}


\end{frontmatter}

% \linenumbers

\section{Introduction}\label{1}

The XFEL uses relativistic electron beams as the medium to generate X-rays, supporting many cutting-edge researches in various fields\cite{huang2021features}. The free-electron laser facility is a device that utilizes relativistic electron beams as the gain medium to amplify the initial electromagnetic field. A normal-conducting linear accelerator is employed to generat,e relativistic electron beams that propagate through undulator systems to produce intense coherent radiation pulses. The FEL pulse comes from the strong and continuous interaction between an electromagnetic wave and the relativistic electron beam in undulator line\cite{yan2021first}. 

The Shanghai soft X-ray Free-electron Laser facility (SXFEL) based on self-amplified spontaneous emission (SASE) and seeded FEL principles is the first fourth-generation light source in China. At present, the SXFEL is under commissioning to provide 2 $\sim$ 10 nm lasing by the end of 2021. There are two undulator lines for the SXFEL user facility. One will be operated with either a cascaded HGHG or EEHG-HGHG mode\cite{yan2021self,feng2019coherent} and named seeded line, while the other will be operated in SASE mode and named SASE line. The seeded line has 18 3-meter-long undulators and includes a structure which containing corrector, quadrupole, BPM, profile and corrector located in the space between every two undulators. The seeded line consists some functional segments such as a seed laser system, three modulators, two radiator sections, and a couple of chicanes serving for laser injection, dispersive section, and fresh bunch delay purposes. The total length of the SASE line is 52 m, including 10 4-meter-long undulators. At front of each undulator, there is a quadrupole with length of 0.1 m, and behind each undulator, there is a BPM with resolution of 1 $\mu$m. There is also a quadrupole and a BPM in the front of the SASE line. In this paper, we perform BBA simulation on SASE line.

The magnetic center of the quadrupole and the electrical center of the BPM will have positional deviations relative to the desired trajectory due to installation errors, natural settlement and other reasons. When the beam enters the undulator line which has offset of quadrupoles, there will be position and direction deviations between the true trajectory and the desired trajectory. As results, the following adverse effects will be produced:

\begin{itemize}
\item The beam can not pass through the magnetic center of the quadrupole, which will produce dispersion effect and then cause the increase of beam emittance.
\item The transverse dipole field generated by the quadrupole deflects the beam trajectory, which cannot meet the requirement of alignment in the undulator part.
\item BPM reflects the position of the beam with error.
\end{itemize}

The SXFEL puts very tight tolerances on the straightness of the trajectory\cite{zeng2019beam} and the emittance of the beam. The offset of the quadrupole can be adjusted by the mover, and the offset of the BPM can be eliminated by processing the readings by the software. How to realize fast and efficient automatic beam alignment on SXFEL is still worth studying.

This trajectory can be achieved by the beam-based alignment (BBA) technique\cite{raubenheimer1991dispersion} which was proposed in 1980s. The BBA technique has achieved great success on LCLS FEL undulator line\cite{emma1999beam}, SACLA undulator line\cite{yamamoto2010beam}, FLASH-I Undulator line\cite{gubeam}, European XFEL SASE1 line\cite{jin2013beam} and PAL-XFEL undulator line\cite{parc2015automation} in 1999, 2011, 2013, 2014 and 2015 respectively. The most widely-used BBA method is the electron beam-based alignment (e-BBA), which mainly includes ballistic alignment (BA)\cite{raubenheimer1999ballistic}, corrector pattern (CP)\cite{aiba2012beam}, Dispersion-Free Steering (DFS)\cite{raubenheimer1991dispersion}. In the DFS method, the singular value decomposition (SVD) method is always used to solve the problem of non-square matrix, in which ``soft constraint'' is used to prevent the emergence of non-convergent solutions\cite{emma1992beam}. Photon beam based alignment (p-BBA) as a complementary tool for e-BBA has been applied on SACLA\cite{tanaka2012undulator} and SDUV-FEL\cite{feng2014coherent}. After e-BBA, p-BBA is used for optimization of undulator magnetic gap, phase match of adjacent undulator segments
or taper of the entire undulator. 

In this paper, we initially introduce two undulator lines of SXFEL. Then we analyze the theory of e-BBA and propose a new automation procedure of e-BBA for the SXFEL SASE line. The two-to-two correction structure is established to change the deviations of position and direction of the beam entering the undulator line with ``K modulation'' method. The DFS method is used to eliminate the offsets of quadrupole and BPM and the ``beam info constraint'' is introduced to replace ``soft constraint''. Ultimately, we give the process and result of simulation.

\section{Theory of BBA} \label{section:2}

\subsection{K Modulation}

The K modulation method\cite{schmidt1993misalignments} is used to measure the relative deviation of the BPM from the quadrupole magnetic center. Under the influence of the transverse dipole field generated by the quadrupole, the beam receives a kick angle and its forward direction is changed when it does not pass through the quadrupole magnetic center. The kick angle is proportional to the focusing intensity $K$ of the quadrupole, so modulating the $K$ of the quadrupole can cause changes of the downstream beam position. At the same time, the kick angle is proportional to the distance between the position where the beam passes through the quadrupole and the quadrupole magnetic center. The change of the downstream beam position can be expressed by\cite{tenenbaum2001developments}:

\begin{equation}
\label{eq1}
{\Delta y} \propto {\Delta K} \cdot {\Delta l}
\end{equation}

\noindent where $\Delta y$ is the change of the beam position measured by downstream BPM, $\Delta K$ is the change of the focusing intensity of quadrupole, $\Delta l$ is the distance between the beam position and the center of quadruple.

After adding a fixed modulation to the $K$ of quadrupole, the downstream beam position change is only related to $\Delta l$. When the beam passes through the center of quadrupole, $\Delta l$ is zero and $\Delta y$ is also zero. At this time, the downstream BPM reading is the offset between the quadrupole and the BPM. It may give inaccurate results if the beam trajectory has a pronounced angle at the quadrupole\cite{marti2020fast}, but this does not affect the judgment of the beam passing through the center of quadrupole.

In this paper, the $K$ of quadrupole is added the ac modulation which is expressed as:

\begin{equation}
\label{eq2}
K = {\Delta K} \cdot cos(2\pi \cdot \frac{t}{T}) + K_{0}
\end{equation}

\noindent where $\Delta K$ is the change range of the focusing intensity, $t$ is the current time, $T$ is the period, and $K_{0}$ is the designed focusing intensity of the quadrupole. 

\subsection{Dispersion-Free Steering}

\begin{figure}[h]
	\centering
	\includegraphics[scale=0.40]{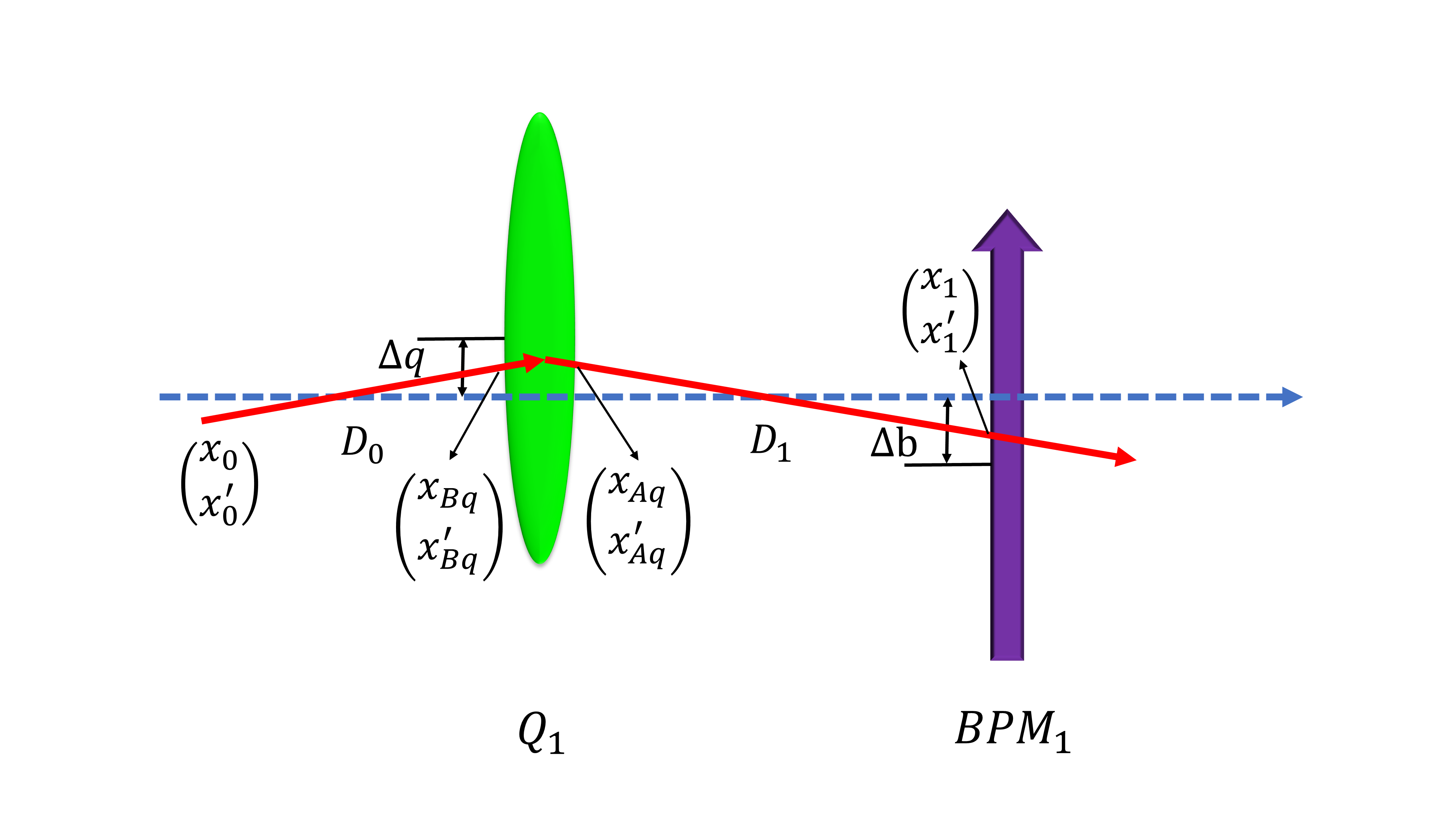}
	\caption{The lattice of the SXFEL SASE line.}
	\label{lattice}
\end{figure}

The lattice of the SXFEL SASE line is shown schematically in Fig.~\ref{lattice}. In BBA progress, the influence of the undulator on the beam trajectory can be ignored when the undulator gap is open. The undulator located between $\rm Q_{1}$ and $\rm BPM_{1}$ and the correctors which are put at undulator ends are not shown in the figure. In Fig.~\ref{lattice}, $x_{0}$($x_{0}^{'}$) is the beam initial position (launch angle), $\Delta q$($\Delta b$) is the offset of the quadrupole (BPM), $x_{Bq}$($x_{Bq}^{'}$) is the beam position (launch angle) before passing through quadrupole, $x_{Aq}$($x_{Aq}^{'}$) is the beam position (launch angle) after passing through quadrupole, and $x_{1}$($x_{1}^{'}$) is the beam position (launch angle) at $\rm BPM_{1}$.

Through the principle of transfer matrix, it is easy to solve $x_{1}$ and $x_{1}^{'}$. However, it is worth noting that when the beam passes through the quadrupole, we need to use the position of the beam relative to the quadrupole magnetic center to perform matrix operations. The beam position $x_{Bq}$ should be replaced by $x_{Bq} - \Delta q$. The position obtained by matrix calculation is also relative to the quadrupole, and then ∆q needs to be added to obtain the real position $x_{Aq}$. The value of $x_{1}$ can be expressed by the Eq.~\ref{eq3}.

\begin{equation}
\label{eq3}
x_{1} = (1+\frac{l_{1}}{f})x_{0}+(l_{0}+l_{1}+\frac{l_{0}l_{1}}{f})x_{0}^{'}-\frac{l_{q}}{f}\Delta q
\end{equation}

\noindent where $l_{0}$, $l_{1}$ and $l_{q}$ are the length of $D_{0}$, $D_{1}$ and $Q_{1}$ respectively. $\frac{1}{f}$ is the $(2,1)$ component of the transfer matrix of quadrupole with thin lens approximation.

Actually, the beam position can only be known by the BPM reading. Due to the existence of BPM offset, the BPM reading is given as:

\begin{equation}
\label{eq4}
x_{read} = x_{1} - \Delta b
\end{equation}

At the same time, the greater the beam energy the smaller the influence of the magnetic field on the beam trajectory. In summary, the BPM reading $x_{read}$ is affected by the initial beam position $x_{0}$, initial launch angle $x_{0}^{'}$, quadrupole offset $\Delta q$, BPM offset $\Delta b$ and beam energy. For the relationship between the BPM readings in the entire SXFEL SASE line and its influencing factors, we can use Ep.~\ref{eq5} to express.

\begin{equation}
\label{eq5}
\begin{aligned}
\begin{pmatrix}
\begin{smallmatrix}
	x_{read1,E_{1}} \\ 
	\vdots \\ 
	x_{readN,E_{1}} \\ 
	x_{read1,E_{2}} \\ 
	\vdots \\ 
	x_{readN,E_{2}} \\
	x_{read1,E_{3}} \\ 
	\vdots \\ 
	x_{readN,E_{3}} 
\end{smallmatrix}
\end{pmatrix} =
\begin{pmatrix}% \setlength{\arraycolsep}{0.05 pt}
\begin{smallmatrix}
	L_{11,E_{1}} & L_{12,E_{1}} & 0 & 0 & 0 & 0 & R_{11,E_{1}} & \cdots & R_{1N,E_{1}} &    \\
	\vdots & \vdots & \vdots & \vdots & \vdots & \vdots & \vdots & \ddots & \vdots &  -I_{NN}\\
	L_{N1,E_{1}} & L_{N2,E_{1}} &  0 & 0 & R_{N1,E_{1}} & R_{NN,E_{1}} &    \\
	0 & 0 & L_{11,E_{2}} & L_{12,E_{2}} & 0 & 0 & R_{11,E_{2}} & \cdots & R_{1N,E_{2}} &    \\
	\vdots & \vdots & \vdots & \vdots & \vdots & \vdots & \vdots & \ddots & \vdots &  -I_{NN}\\
	0 & 0 & L_{N1,E_{2}} & L_{N2,E_{2}} & 0 & 0 & R_{N1,E_{2}} & \cdots & R_{NN,E_{2}} &    \\
	0 & 0 & 0 & 0 & L_{11,E_{3}} & L_{12,E_{3}} & R_{11,E_{3}} & \cdots & R_{1N,E_{3}} &    \\
	\vdots & \vdots & \vdots & \vdots & \vdots & \vdots & \vdots & \ddots & \vdots &  -I_{NN}\\
	0 & 0 & 0 & 0 & L_{N1,E_{3}} & L_{N2,E_{3}} & R_{N1,E_{3}} & \cdots & R_{NN,E_{3}} &
\end{smallmatrix}    
\end{pmatrix}
\begin{pmatrix}
\begin{smallmatrix}
	x_{0,E_{1}} \\ 
	x_{0,E_{1}}^{'} \\
	x_{0,E_{2}} \\ 
	x_{0,E_{2}}^{'} \\
	x_{0,E_{3}} \\ 
	x_{0,E_{3}}^{'} \\
	\Delta q_{1}  \\
	\vdots   \\
	\Delta q_{N}\\
	\Delta b_{1} \\ 
	\vdots \\ 
	\Delta b_{N} 
\end{smallmatrix}
\end{pmatrix}
\end{aligned}
\end{equation}

\noindent where $x_{read1,E_{1}}$ is the beam position reading at the 1st BPM when the electron beam energy is $E_{1}$. $N$ is the number of the quadrupoles and BPMs. The calculation method of other elements in the matrix can be found in Ref.\cite{parc2015automation}.

The Eq.~\ref{eq1} can be rewritten as:

\begin{equation}
\label{eq6}
\boldsymbol{X} = \boldsymbol{RO}
\end{equation}

\noindent where $\boldsymbol{R}$ is the full response matrix of $3N\times(2N+6)$ size.

According to the knowledge of linear algebra, there are multiple matrices $\boldsymbol{O}$ that make the equation true. In order to obtain a realistic solution, ``soft constraint'' is usually added to Eq.~\ref{eq1}. A new constraint which named ``beam info constraint'' is proposed in this paper. It requires that the initial position and initial launch angle of the beam remain unchanged when obtaining BPM readings under two different energies. This constraint is expressed as: 

\begin{equation}
\label{eq7}
\left\{
\begin{array}{lr}
x_{0,E_{1}} - x_{0,E_{2}}=0, & \\
x_{0,E_{1}}^{'} - x_{0,E_{2}}^{'}=0, & \\
x_{0,E_{1}} - x_{0,E_{3}}=0,  & \\  
x_{0,E_{1}}^{'} - x_{0,E_{3}}^{'}=0. &\\
\end{array}
\right.
\end{equation}

After adding constraint to the Eq.~\ref{eq1}, the size of $\boldsymbol{R}$ becomes $(3N+4)\times(2N+6)$. Since $\boldsymbol{R}$  is not a square matrix, SVD needs to be used to obtain the generalized inverse matrix:

\begin{equation}
\label{eq8}
\boldsymbol{R} = \boldsymbol{USV^{T}}
\end{equation}

The matrix $\boldsymbol{O}$ can be written as:
\begin{equation}
\label{eq9}
\boldsymbol{O} = \boldsymbol{R^{-1}X} =  \boldsymbol{VS^{-1}U^{T}X}
\end{equation}

If the initial beam position and the launch angle are ignored, the Eq.~\ref{eq1} can be written as:

\begin{equation}
\label{eq10}
\begin{pmatrix}
\begin{smallmatrix}
	x_{read1,E_{1}} \\ 
	\vdots \\ 
	x_{readN,E_{1}} \\ 
	x_{read1,E_{2}} \\ 
	\vdots \\ 
	x_{readN,E_{2}} \\
	x_{read1,E_{3}} \\ 
	\vdots \\ 
	x_{readN,E_{3}}
\end{smallmatrix}
\end{pmatrix}  = 
\begin{pmatrix}
\begin{smallmatrix}
	R_{11,E_{1}} & \cdots & R_{1N,E_{1}} &    \\
	\vdots & \ddots & \vdots &  -I_{NN}\\
	R_{N1,E_{1}} & R_{NN,E_{1}} &    \\
	R_{11,E_{2}} & \cdots & R_{1N,E_{2}} &    \\
	\vdots & \ddots & \vdots &  -I_{NN}\\
	R_{N1,E_{2}} & R_{NN,E_{2}} &    \\
	R_{11,E_{3}} & \cdots & R_{1N,E_{3}} &    \\
	\vdots & \ddots & \vdots &  -I_{NN}\\
	R_{N1,E_{3}} & \cdots & R_{NN,E_{3}} &    \\
\end{smallmatrix}
\end{pmatrix}
\begin{pmatrix}
\begin{smallmatrix}
	\Delta q_{1}  \\
	\vdots   \\
	\Delta q_{N}\\
	\Delta b_{1} \\ 
	\vdots \\ 
	\Delta b_{N}\\
\end{smallmatrix}
 \end{pmatrix}
\end{equation}

The Eq.~\ref{eq10} can be rewritten as:

\begin{equation}
\label{eq11}
\boldsymbol{X} = \boldsymbol{rO}
\end{equation}

\noindent where $\boldsymbol{r}$ is the response matrix of $3N\times2N$ size. When two energies are selected, $\boldsymbol{r}$ can become a square matrix, but for calculation accuracy, three energies are still used for calculation.

The offsets of quadrupole and BPM from the straight-line orbit determined by beam initial position and launch angle can be obtained by solving the Eq.~\ref{eq10} by SVD method. According to this solution, the quadrupole center can be moved to the beam orbit, that is, the beam passes through the quadrupole magnetic center.

\subsection{Alignment steps}
\begin{itemize}
\item[] Step 1: Using the full response matrix to solve the offsets of quadrupole and BPM from the ideal orbit. This step can be named ``ideal orbit-based alignment (IOBA)'' in this paper.
\item[] Step 2: Using two-to-two correction to change the initial position and launch angle of beam. This step can be named ``beam info alignment (BIA)'' in this paper.
\item[] Step 3: Using the response matrix to solve the offsets of quadrupole and BPM from the straight-line orbit determined by beam initial position and launch angle. This step can be named ``beam orbit-based alignment (BOBA)'' in this paper.
\end{itemize}

\section{Simulation result}

In order to simulate the beam transmission process, a simple simulation program is written using Python and PyQt. The program can construct a system model based on the transmission matrix of the element and calculate the beam position. Compared with the traditional simulation software ELEGANT\cite{borland2000flexible} and GENESIS\cite{reiche1999genesis}, the simulation program used in this paper has a visual interface. Through the simulation program, the offsets of quadrupole and BPM can be changed and the real-time changes of the beam orbit can be displayed on the interface.

\begin{figure}[!ht]
	\centering
	\includegraphics[scale=0.25]{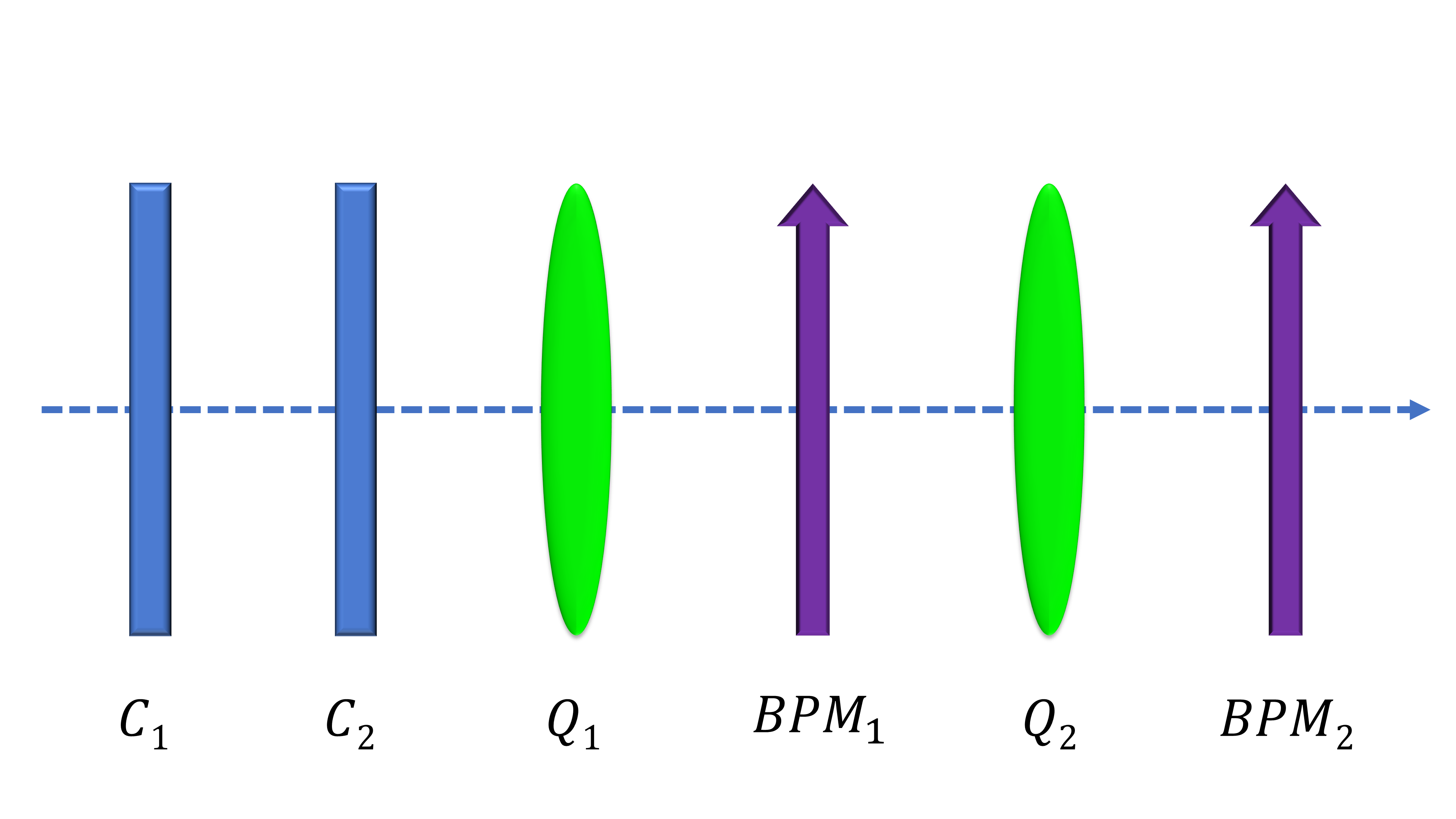}
	\caption{The lattice of the SXFEL SASE line.}
	\label{2to2}
\end{figure}

In the actual BBA, the gap of the undulator is enlarged, and the influence on the beam orbit is ignored. Therefore, the undulator is used as a drift section for calculation in the simulation. In the second step of the alignment method, a two-to-two correction structure is needed to change the position and launch angle of the beam entering the undulator section, so we add the two-to-two correction structure as shown in the Fig.~\ref{2to2} in front of the SASE line. The kick angles that the two correctors ultimately need to provide are inversely proportional to the distance of the drift section between the components. In this simulation, the length of each drift in the two-to-two correction structure is set to 0.5 m. The error range of the various components we used in the simulation is shown in the Tab.~\ref{table1}.

\begin{figure}[!ht]
	\centering
	\includegraphics[scale=0.40]{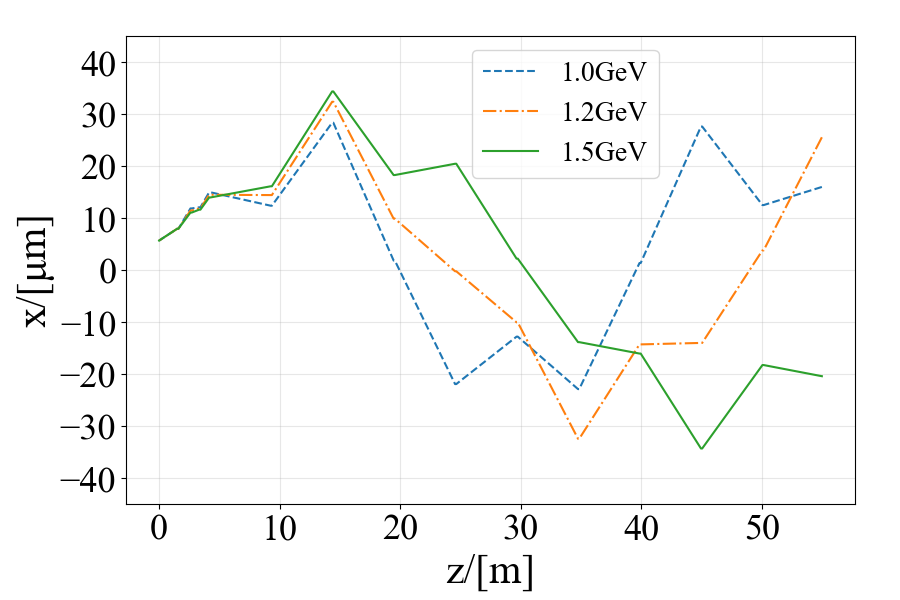}
	\caption{Beam trajectory with no position offsets of quadrupole and BPM.}
	\label{orbit_with_offset}
\end{figure}

\begin{figure}[!ht]
	\centering
	\includegraphics[scale=0.40]{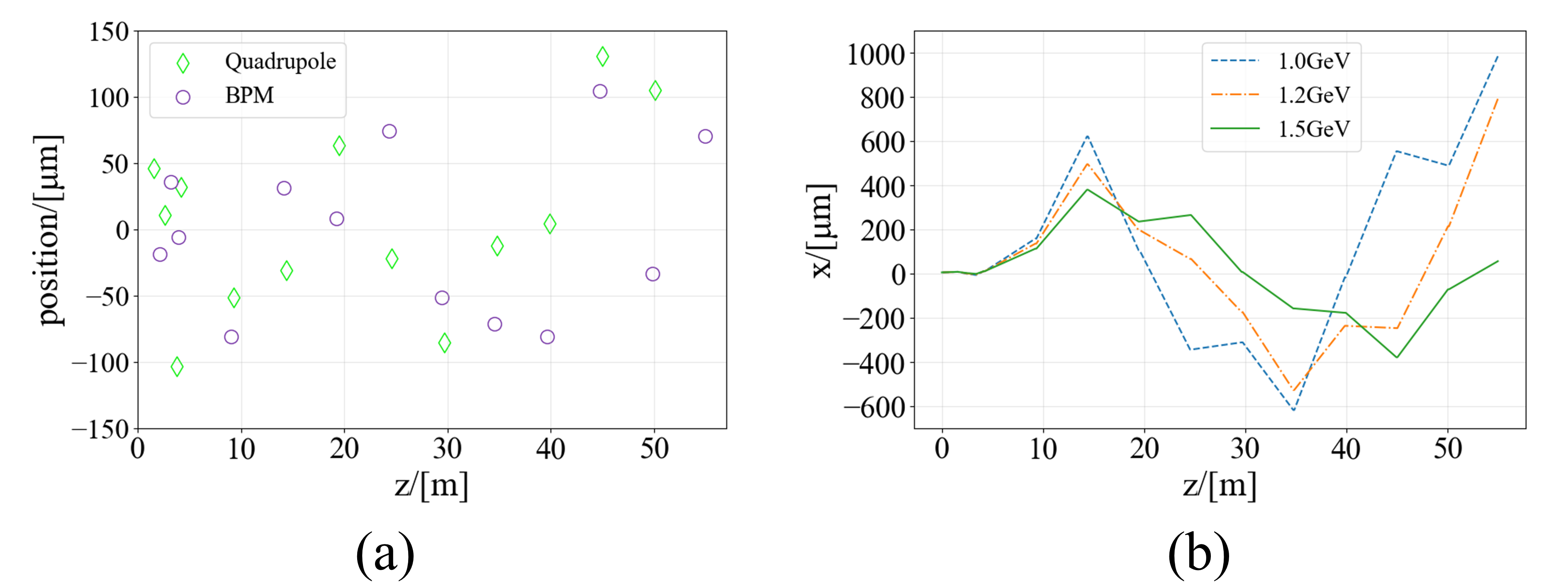}
	\caption{(a) is the position of quadrupoles and BPMs after adding offsets. (b) is the beam orbit after adding offsets to quadrupole and BPM.}
	\label{initial}
\end{figure}

In this simulation, the initial electron beam position was set to 5.66 $\mu$m and the launch angle of the electron beam to the design orbit was set to 1.52 $\mu$rad. We select beam energies of 1.0 GeV, 1.2 GeV and 1.5 GeV in the simulation for the SXFEL SASE line. The initial beam parameters are randomly generated according to the gaussian distribution. The beam trajectory in Fig.~\ref{orbit_with_offset} shows that the electron beam with energy of 1 GeV, 1.2 GeV and 1.5 GeV can be max about 30 $\mu$m off from the ideal orbit. Then quadrupoles and BPMs are added with random offset as shown in Fig.~\ref{initial}(a). The rms value and Standard Deviation ($\sigma$) of quadrupole positions are 66.54 $\mu$m and 59.50 $\mu$m. For BPM, those values are 66.20 $\mu$m and 59.49 $\mu$m. The beam orbits are shown in Fig.~\ref{initial}(b). The initial beam orbits under energy of 1 GeV, 1.2 GeV and 1.5 GeV have rms of 415.43 $\mu$m, 300.76 $\mu$m and 207.43 $\mu$m respectively, which cannot meet the requirement of X-ray laser generation in the undulator line.

\begin{table}[!ht]
\setlength{\belowcaptionskip}{8pt}
   \centering
   \caption{Errors Used in the BBA Simulation}
   \begin{tabular}{lccc}
       \toprule
       \textbf{Errors (rms)}   & \textbf{Value} & \textbf{Unit}\\
       \midrule
          Quadrupole offset  & 100 & $\mu$m  \\ %[3pt]
          BPM offset  & 100 & $\mu$m  \\ %[3pt]
          BPM resolution  & 1 & $\mu$m  \\ %[3pt]
          Beam launch position  & 10 & $\mu$m  \\ %[3pt]2
          Beam launch angle  & 10 & $\mu$rad  \\ %[3pt]
          
       \bottomrule
   \end{tabular}
   \label{table1}
\end{table}

From the comparison before and after adding the offset for quadrupoles and BPMs, it can be seen that the existence of quadrupole offset greatly increases the beam orbit deviation. In order to get lasing, the offset of quadrupole must be eliminated, and then the simulation was carried out according to the alignment method mentioned in Section.~\ref{section:2}.

\subsection{Step 1: IOBA}
\begin{figure}[!ht]
	\centering
	\includegraphics[scale=0.40]{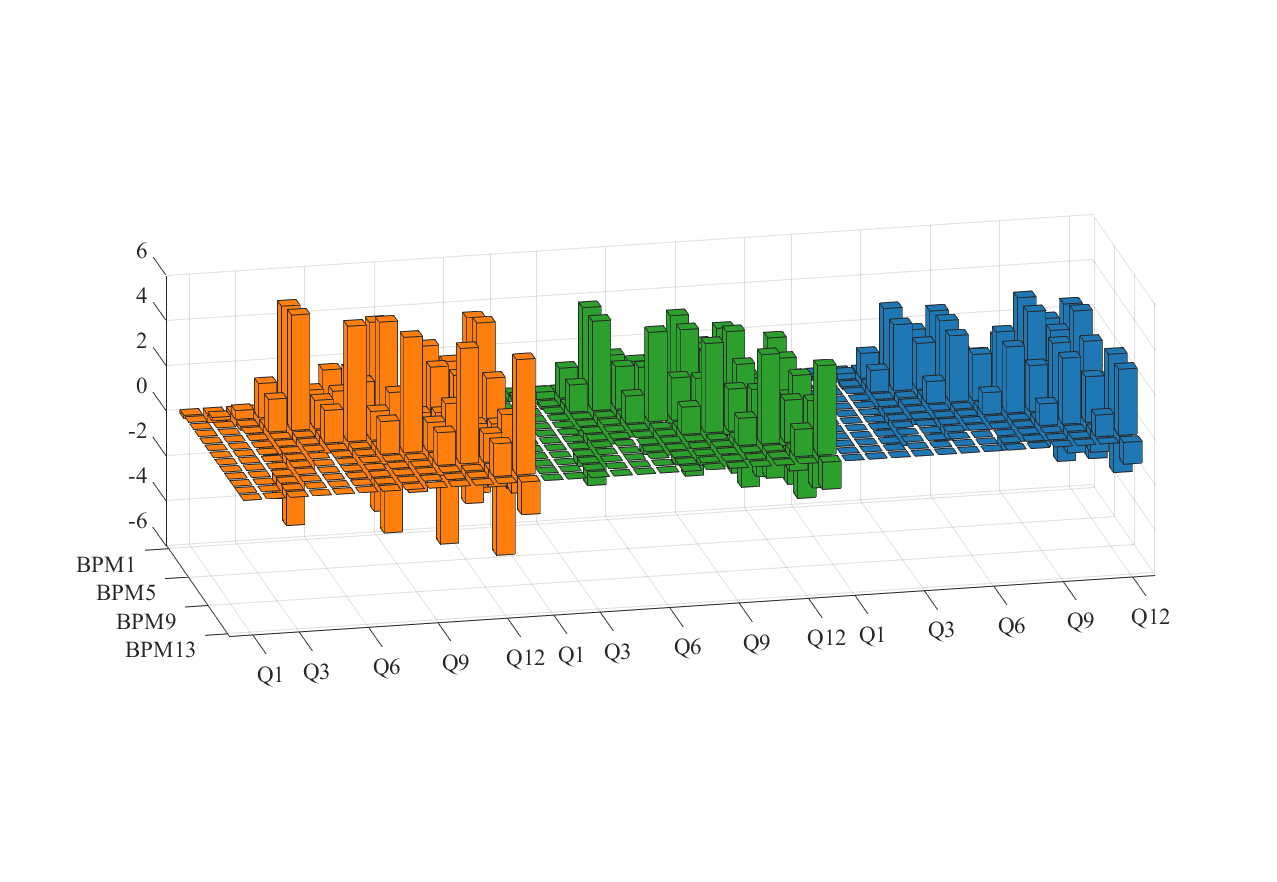}
	\caption{The response matrix for 1.0 GeV (orange), 1.2 GeV (green) and 1.5 GeV (blue).}
	\label{RM}
\end{figure}

IOBA is used to adjust the position of quadrupoles from about 60 $\mu$m rms to 10 $\mu$m rms. Firstly, we need to solve the full response matrix of the entire undulator line including the two-to-two correction structure. In the simulation, the model of the undulator entire line is constructed according to the principle of transmission matrix. Fig.~\ref{RM} illustrates the response matrix at different energies.

\begin{figure}[!ht]
	\centering
	\includegraphics[scale=0.40]{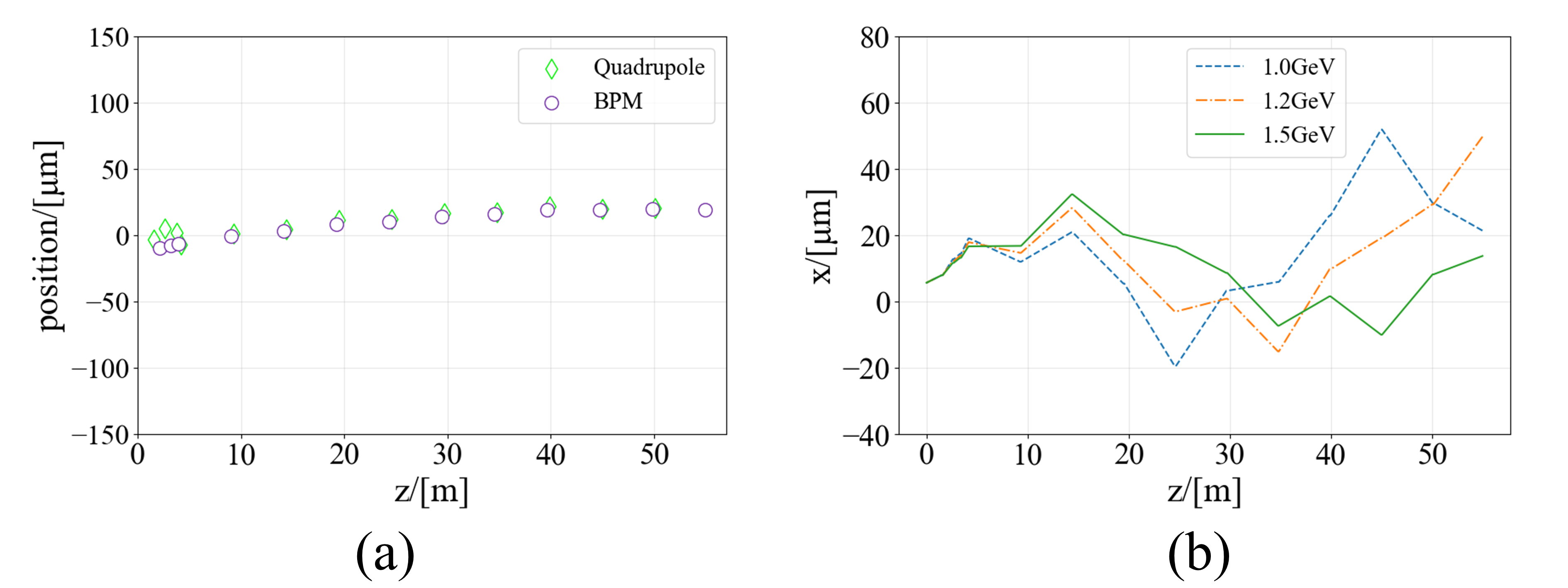}
	\caption{(a) and (b) are the position of quadrupoles and BPMs and beam orbit after IOBA.}
	\label{IOBA}
\end{figure}

After three iterations of IOBA, the randomly distributed quadrupoles and BPMs are aligned with a line shape as shown in Fig.~\ref{IOBA}(a). After IOBA, the rms and standard deviation ($\sigma$) of quadrupole positions become 13.28 $\mu$m and 13.49 $\mu$m. For BPM, those values become 9.23 $\mu$m and 10.73 $\mu$m. Compared with the initial position, the rms of quadrupole position and BPM position decreased by 80.04\% and 86.06\%, and the standard deviation decreased by 77.33\% and 81.96\%. It shows that the IOBA can well reduce the offsets of quadrupole and BPM and greatly improve the horizontal alignment of their centers. Fig.~\ref{IOBA}(b) shows the beam orbit after IOBA. After moving the quadrupole, the rms of the beam orbit at 1.0 GeV, 1.2 GeV and 1.5GeV is 22.86 $\mu$m, 19.23 $\mu$m and 15.53 $\mu$m, respectively. The beam orbit can be restored to the level when there is no position offsets of the quadrupole and BPM. The $\sigma$ of the beam orbit at 1.0 GeV, 1.2 GeV and 1.5GeV drops from 403.25 $\mu$m to 16.77 $\mu$m, from 299.66 $\mu$m to 14.25 $\mu$m and from 206.12 $\mu$m to 11.45 $\mu$m, respectively. The simulation results show that the max offsets of electron beam orbits are improved from 100 $\mu$m level to 10 $\mu$m level for each beam energy and the $\sigma$ of electron beam orbits are improved from 100 $\mu$m level to 1 $\mu$m level. But the beam orbits are not straight enough.

\subsection{Step 2: BIA}
\begin{figure}[!ht]
	\centering
	\includegraphics[scale=0.40]{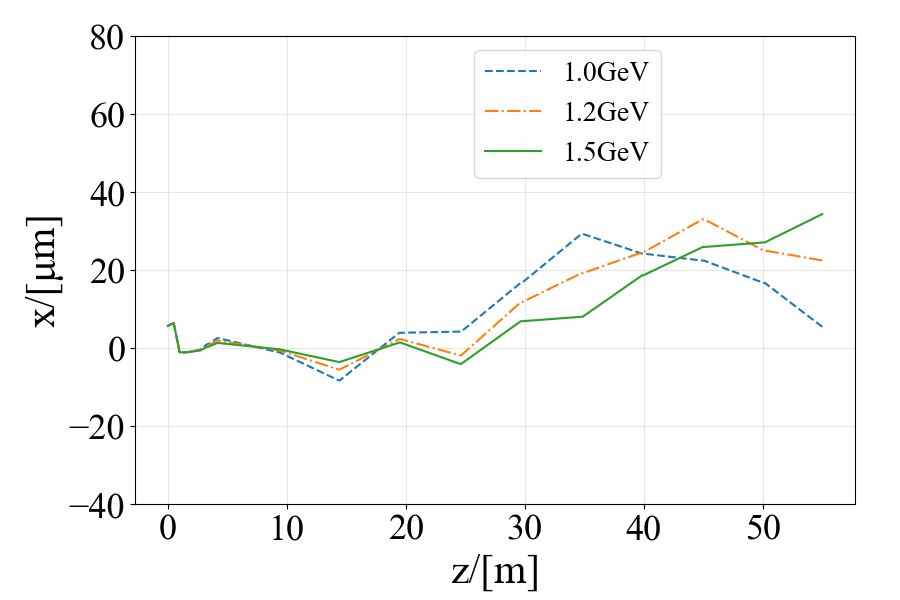}
	\caption{Beam orbit under different energies after BIA.}
	\label{BIA}
\end{figure}

In this step, K modulation is applied to the two quadrupoles in two-to-two correction structure to determine whether the beam passes through the quadrupole magnetic center. We adjust the kick angle $\theta _{1}$ and $\theta _{2}$ provided by the two correctors to make the beam roughly pass through the center of the two quadrupoles and ensure that the sum of $\theta _{1}$, $\theta _{2}$ and the initial beam launch angle $x_{0}^{'}$ is zero. In this way, we can ensure that the beam position does not deviate too much from the design orbit and the beam launch angle is close to zero before the beam entering the undulator line. After adjustment using the automation program, when $\theta _{1}$ is -16.58 $\mu$rad and $\theta _{2}$ is 15.07 $\mu$rad, the beam passes through the center of two quadrupoles and the launch angle is close to zero, as shown in Fig.~\ref{BIA}. We find that the deviation of the beam orbit has also been improved, but the most important thing is that the beam launch angle before entering undulator line is changed.

\subsection{Step 3: BOBA}

We use BOBA to correct the offsets of quadrupole and BPM again. We take the structure behind the corrector $C_{2}$ in the two-to-two correction structure as the new overall structure for the corresponding matrix calculation. The main purpose is using the beam position and angle adjusted by the two correctors as the initial position and angle of the beam entering the undulator line. The offsets of quadrupole and BPM with respect to the beam straight-line orbit determined by the initial beam position and beam angle are calculated by the response matrix which does not consider the initial information of the beam.

\begin{figure}[!ht]
	\centering
	\includegraphics[scale=0.40]{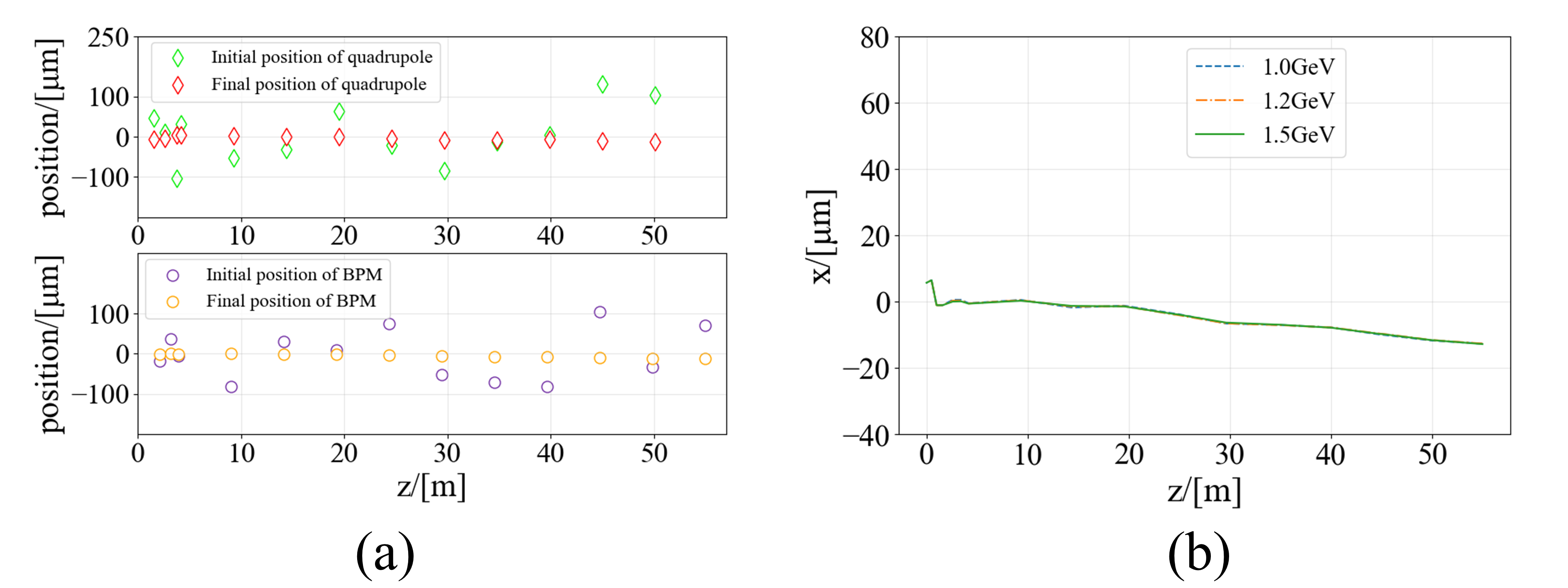}
	\caption{(a) and (b) are the position of quadrupoles and BPMs and beam orbit after BOBA.}
	\label{BOBA}
\end{figure}

Fig.~\ref{BOBA}(a) shows the position of quadrupole and BPM after BOBA. The rms and $\sigma$ of quadrupole positions become 6.54 $\mu$m and 5.49 $\mu$m. For BPM, those values become 6.63 $\mu$m and 4.58 $\mu$m. Compared with the beam orbits obtained by IOBA, the rms of the quadrupole and BPM positions have both increased slightly, but the $\sigma$ has been very small, that is, the centers of these devices have been basically on the horizontal line. After adjustment, the beam orbits under three energies basically coincide to eliminate the dispersion effect, and the beam orbits in undulator line are very collimated, that is, the beam passes through the magnetic center of quadrupoles. The rms of the beam orbit is 6.63 $\mu$m and the $\sigma$ is 4.20 $\mu$m, which is better than the beam orbit when the quadrupoles don't have offset. The final beam orbits after BOBA are shown in Fig.~\ref{BOBA}(b).

\begin{figure}[!ht]
	\centering
	\includegraphics[scale=0.40]{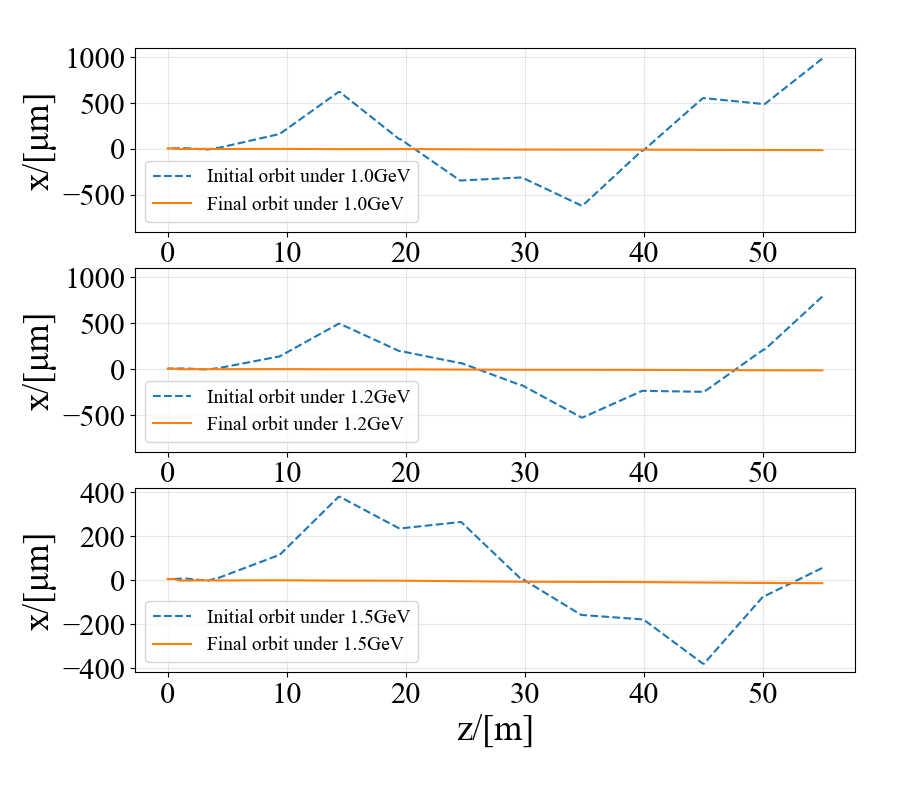}
	\caption{The beam orbits before and after BBA.}
	\label{FOIO}
\end{figure}

Fig.~\ref{FOIO} and Tab.~\ref{table2} respectively show the schematic diagram and detailed parameters of the electron beam trajectory before and after BBA. The result indicates a significant improvement of the beam orbits.

\begin{figure}[!ht]
	\centering
	\includegraphics[scale=0.40]{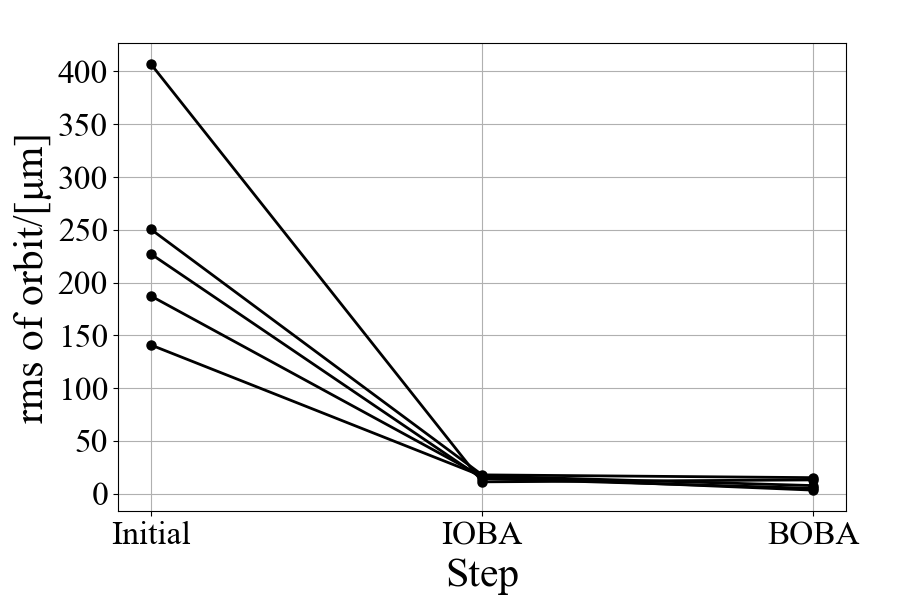}
	\caption{Results with 5 random seeds. The rms of orbit after alignment.}
	\label{verification}
\end{figure}

The applicability of the method is verified by using 5 sets of randomly generated beam initial position, beam initial launch angle, offsets of quadrupole and BPM. Fig.~\ref{verification} shows the change of beam orbit rms under 1.5 GeV and we can see that rms of different beam orbits are around 10 $\mu$m. This shows that after three steps alignment operation, a better beam trajectory can be obtained.

\begin{table}[!ht]
\setlength{\belowcaptionskip}{8pt}
   \centering
   \caption{Parameters of beam orbits before and after BBA}
   \begin{tabular}{lcccc}
       \toprule
       \multicolumn{2}{c}{\textbf{Description}}  & \textbf{Initial($\mu$m)} & \textbf{Final($\mu$m)}\\
       \midrule
          \multirow{2}*{1.0 GeV} & rms of orbit  & 415.43 & 6.69  \\ %[3pt]
          ~ & $\sigma$ of orbit  & 403.25 & 4.29  \\ %[3pt]
          \multirow{2}*{1.2 GeV} & rms of orbit  & 300.76 & 6.64  \\ %[3pt]
          ~ & $\sigma$ of orbit  & 299.66 & 4.20  \\ %[3pt]
          \multirow{2}*{1.5 GeV} & rms of orbit  & 207.43 & 6.63  \\ %[3pt]
          ~ & $\sigma$ of orbit  & 206.12 & 4.20  \\ %[3pt]
          \multirow{2}*{Desipersion $\eta$} & max  & 2795.80 & 1.44  \\ %[3pt]
          ~ & rms  & 1360.06 & 0.65  \\ %[3pt]

       \bottomrule
   \end{tabular}
   \label{table2}
\end{table}

\section{Summary}
In this paper, we proposed putting a two-to-two correction structure before the undulator line and using three steps to collimate the beam trajectory. In order to verify the accuracy of the method, we conducted a simulation. In the first step, we proposed to add a new constraint named ``beam info constraint'' to the SVD calculation of the full response matrix. Finding by the simulation which applying this constraint that the offsets of quadrupole and BPM can be corrected to about 10 $\mu$m. In the second step, we applied the K modulation to the quadrupoles which are in the two-to-two correction structure to adjust the beam passing through the quadrupole magnetic centers. The main purpose is to change the position and launch angle of the beam entering the undulator line. In the third step, we used the response matrix to calculate the offsets of quadrupole and BPM relative to the straight-line orbit determined by the position and angle of the beam entering the undulator line and adjust the positions of quadrupole and BPM. After multiple sets of simulation verifications, a highly collimated beam trajectory with rms and $\sigma$ at the level of 1 $\mu$m can be obtained. For further optimization, the method will cooperate with p-BBA to introduce to SXFEL undulator lines and has great helpful for FEL lasing.

%%%%%%%%%%%%%%%%%%%%%%%%%%%截止%%%%%%%%%%%%%%%%%%%%%%%%%%%%%%%%%%%%%

\section{Acknowledgement}

This work was supported by the National Key Research and Development Program of China (2018YFE0103100), the National Natural Science Foundation of China (12125508, 11935020) and Program of Shanghai Academic/Technology Research Leader (21XD1404100). The authors would like to thank Hanxiang Yang, Li Zeng and Jiawei Yan for fruitful discussions.

%\section*{References}

\bibliography{main}

\end{document}